\documentclass[showkeys,showpacs,prd]{revtex4}
\usepackage{amsmath}
\usepackage{amssymb}
\usepackage{graphicx}
\usepackage{color}

\begin{document}

\title{
Calculation of gluon contribution to the proton spin by using the non-perturbative quantization \`{a} la Heisenberg
}

\author{
Vladimir Dzhunushaliev
}
\email{v.dzhunushaliev@gmail.com}
\affiliation{
	Dept. Theor. and Nucl. Phys., KazNU, Almaty, 050040, Kazakhstan \\
	IETP, Al-Farabi KazNU, Almaty, 050040, Kazakhstan \\
	Institute of Systems Science,
	Durban University of Technology, P. O. Box 1334, Durban 4000, South Africa
}

\begin{abstract}
The contribution of crossed gluon fields in flux tubes connecting quarks to the proton spin is calculated. The calculations are performed following non-perturbative Heisenberg's quantization technique. In our approach a proton is considered as consisting of three quarks connected by three flux tubes. The flux tubes contain colour longitudinal electric and transversal electric and magnetic fields. The transversal fields causes the appearance of the angular momentum density. The dimensionless relation between the angular momentum and the mass of the gluon fields is obtained. The contribution to proton spin from rotating quarks and flux tubes connecting quarks is estimated. Simple numerical relation between the proton mass, the speed of light and the proton radius, which is of the same order as the Planck constant, is discussed.
\end{abstract}

\date{\today}

\maketitle

\section{Introduction}

The spin structure of the proton is one of the most challenging problems in modern physics. The experimental results of the European Muon Collaboration showed that only a small fraction of the proton spin is carried by the quark spin \cite{Ashman:1987hv,Ashman:1989ig}. The proton spin can be split as 
\begin{equation}
	\frac{1}{2} = \frac{1}{2} \Delta \Sigma + \Delta G + 
	L_q + L_G 
\label{i-10}
\end{equation}
where $\Delta \Sigma$ is a quark spin contribution, $\Delta G$ is a gluonic contribution, $L_q$ and $L_G$ are quark and gluon angular momentum contributions. Measurements of the quarks contribution to the proton spin \cite{Ageev:2005gh} - \cite{Alekseev:2010ub} show that it is $\approx 30 \%$. Further studies showed that up to $30 \%$ the proton spin can arise due to the spin of gluons \cite{Adare:2014hsq} \cite{Adamczyk:2014ozi}. Some portion of the proton spin can come from the orbital angular momentum of quarks and anti-quarks \cite{Myhrer:2007cf}, see also the reviews  \cite{Kuhn:2008sy}-\cite{Bass:2004xa}. Theoretical investigations of the proton spin structure are also performed on the lattice, see for example \cite{QCDSF:2011aa}. In Ref. \cite{Ji:1995cu} the general procedure for calculating the quark and gluon helicity contributions $\Delta \Sigma$, $\Delta G$, and the quark and gluon orbital angular momentum contributions $L_q$, $L_G$ to the proton spin is presented. It is possible that $L_G$ is a purely nonperturbative effect and its calculation is closely related to the resolution of the confinement problem. In this note we are guided by this idea and show that the contribution of $L_G$ to the proton spin is connected with the presence of flux tubes between quarks in the proton.

In this letter we use the non-perturbative methods \`{a} la Heisenberg \cite{heis} to investigate the gluon field contribution to the proton spin.
The main idea of our approach is that flux tubes between quarks in a proton have the angular momentum density coming from colour electric and magnetic fields. The flux tube has longitudinal electric field and transversal electric and magnetic fields. The transversal fields have different colour indices and, at first glance, cannot give us the angular momentum density. But if the proton quantum state has a quantum interplay between states with different transversal fields in the flux tube we will have transversal electric and magnetic fields with the same colour index. In this case we can calculate the angular momentum coming from the gluon fields which are in the flux tube. In this connection we have to note that in Ref. \cite{Kochelev:2005vd} the authors have shown that some properties of a glueball may explain the gluonic contribution to the proton spin. 

\section{Main idea}
\label{mIdea}

In this section we would like to present the main idea of calculation of the gluonic contribution to the proton spin. Before proceeding with the numerical calculations, we want to understand qualitatively how the gluon field makes the contribution to the proton spin. Let us consider the following numerical relation between the fundamental constants and the proton parameters
\begin{equation}
	m_p c r_p = 5 \cdot {10}^{-27} \text{g} \cdot \text{cm}^2 \cdot \text{s}^{-1}
	\approx 4.75 \cdot \hbar.
\label{2-10}
\end{equation}
Here $m_p$ is the proton mass and $r_p$ -- the proton radius. The left-hand side of \eqref{2-10} is the angular momentum of something rotating with the speed of light around the proton centre. We know that: (a) the proton mass 
has a significant contribution from gluon fields, (b) a gluon  has zero rest mass and moves with the speed of light. This leads us to a thought that the angular moment of the gluon field has to make a significant contribution to the proton spin. This can happen only if the gluon fields form a kind of ordered structure. For example, it can be flux tubes between quarks. This is the main idea suggested in the letter.

\begin{figure}[h!]
\begin{center}
	\fbox{
		\includegraphics[width=.4\linewidth]{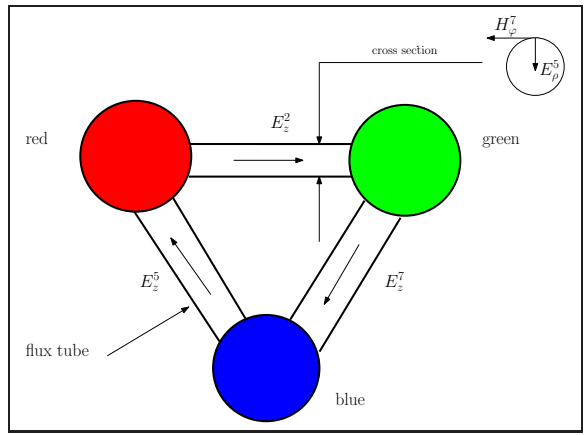}
	}
\end{center}
	\caption{Proton model with three quarks connected by flux tubes. 
		}
\label{protonModel1}
\end{figure}

We use the following model of a proton: three quarks are connected by flux tubes, see Fig.\ref{protonModel1}. In Ref. \cite{Dzhunushaliev:2016svj} it is shown that the flux tube stretched between two infinitely separated quarks has the longitudinal colour electric field $E^3_z$ and two transversal fields: the radial colour electric field $E^1_\rho$ and the azimuthal colour magnetic field $H^2_\varphi$. It is well known  from electrodynamics that electric and magnetic fields may have the angular momentum density
$\vec M = \frac{1}{4 \pi c} \left[
	\vec r \times \left[ \vec E \times \vec H \right]
\right]
$. The generalization for non-Abelian gauge fields is obvious:
$\vec M = \frac{1}{4 \pi c} \left[
	\vec r \times \left[ \vec E^a \times \vec H^a \right]
\right]
$, where $a$ is the colour index. A direct application of these formulae gives us a zero angular momentum density since the colour indices of $\vec E^a$ and $\vec H^b$
for every flux tube are different, $b \neq a$. But following to Section \ref{inftube} (Ansatz \eqref{3-90} - \eqref{3-97}) it is easy to see that having the solution with $A^2_t, A^5_z$ gauge potentials we can exchange either $2 \leftrightarrow 5$ or $t \leftrightarrow z$ indexes and obtain the solution with $A^2_z, A^5_t$ gauge potentials. It allows us to say that in quantum reality we will have a quantum state $\left. \left. \right| p \right\rangle$ where both solutions are realized. In that case 
$
\left\langle p \left|
	\left[
	\vec r \times \left[
		\hat {\vec E}^a \times \hat {\vec H}^a
	\right]
	\right] 	
\right| p \right\rangle \neq 0
$ and then the total angular momentum of a proton has a contribution from the gluon fields (here $\left. \left. \right| p \right\rangle$ is the quantum state of a proton).

\section{Infinite flux tube}
\label{inftube}

In this section we want to obtain an infinite flux tube solution using the non-perturbative quantisation approach
\`{a} la Heisenberg presented in Ref. \cite{Dzhunushaliev:2016svj}. We start with the two-equation approximation obtained in Ref. \cite{Dzhunushaliev:2016svj} and applied for the flux tube.
The set of equations describing such a situation is
\begin{eqnarray}
	\tilde D_\nu F^{a \mu \nu} - \left[
		\left( m^2 \right)^{ab \mu \nu} -
	\left( \mu^2 \right)^{ab \mu \nu}
	\right] A^b_\nu &=& 0 ,
\label{3-10}\\
	\Box \phi - \left( m^2_\phi \right)^{ab \mu \nu} A^a_\nu A^b_\mu \phi -
	\lambda \phi \left( M^2 - \phi^2  \right) &=& 0,
\label{3-20}
\end{eqnarray}
where
\begin{eqnarray}
	\left( m^2 \right)^{ab \mu\nu} &=& - g^2 \left[
		f^{abc} f^{cpq} G^{pq \mu\nu} -
		f^{amn} f^{bnp} \left(
			\eta^{\mu \nu} G^{mp \phantom{\alpha} \alpha}_{\phantom{mn} \alpha} -
			G^{mp \nu \mu}
		\right)
	\right] ,
\label{3-30} \\
	\left( \mu^2 \right)^{ab \mu \nu} &=& - g^2 \left(
	f^{abc} f^{cde} G^{de \mu \nu} +
	\eta^{\mu \nu} f^{adc} f^{cbe} G^{de \phantom{\alpha} \alpha}_{\phantom{de} \alpha} +
	f^{aec} f^{cdb} G^{ed \nu \mu}
	\right) ,
\label{3-40} \\
	\left( m^2_\phi \right)^{ab \mu \nu} &=&
	g^2 f^{amn} f^{bnp} \frac{
		G^{mp \mu \nu} - \eta^{\mu \nu} G^{mp \alpha}_{\phantom{mp \alpha} \alpha}
	} {G^{mm \alpha}_{\phantom{mm\alpha} \alpha}} .
\label{3-60}
\end{eqnarray}
2-point Green functions for the gauge fields $\delta \hat A^a_\mu \in SU(2) \times U(1)$ and for the coset  $\hat A^m_\mu \in SU(3) / SU(2) \times U(1)$ are defined as
\begin{eqnarray}
	G^{mn \mu \nu}(y,x) &=& \left\langle
		\hat A^{m \mu}(y) \hat A^{n \nu}(x)
	\right\rangle , \quad
	\left\langle \hat A^{m \mu} \right\rangle = 0,
\label{3-70}\\
	G^{ab \mu \nu}(y,x) &=& \left\langle
		\delta \hat A^{a \mu}(y) \delta \hat A^{b \nu}(x)
	\right\rangle , \quad
	\hat A^{a \mu} = \left\langle \hat A^{a \mu} \right\rangle +
	i \delta \hat A^{a \mu} ,
\label{3-80}
\end{eqnarray}
where
$F^a_{\mu \nu} = \partial_\mu A^a_\nu - \partial_\nu A^a_\mu + g f^{abc} A^b_\mu A^c_\nu$
is the field strength; $b,c,d = 2,5,7$ are the SU(2) colour indices; $g$ is the coupling constant; $f^{bcd}$ are the structure constants for the SU(3) gauge group and $g$ is the coupling constant.
The set of equations \eqref{3-10}  and \eqref{3-20} describes decomposition of SU(3) degrees of freedom into two groups: the first one describes $SU(2) \in SU(3)$ degrees of freedom and $\phi$
is a gluon condensate describing an average dispersion of quantum fluctuations in the coset space $SU(3) / SU(2)$.

We seek a cylindrically symmetric solution of the set of equations  \eqref{3-10} and \eqref{3-20} in the subgroup $SU(2) \in SU(3)$ spanned on $\lambda^{2,5,7}$ (in Ref. \cite{Dzhunushaliev:2016svj}
we found the solution in the $SU(2)$ subgroup spanned on $\lambda^{1,2,3}$):
\begin{equation}
	A^2_t(\rho) = \frac{f(\rho)}{g} ; \quad
	A^5_z(\rho) = \frac{v(\rho)}{g} ; \quad
	\phi(\rho) = \frac{\phi(\rho)}{g}.
\label{3-90}
\end{equation}
Here we use the cylindrical coordinate system $z, \rho, \varphi$. The corresponding colour electric and magnetic fields are then
\begin{eqnarray}
	E^7_z(\rho) &=& F^3_{tz} = \frac{f(\rho) v(\rho)}{g}  ,
\label{3-93}\\
	E^2_\rho(\rho) &=& F^2_{t \rho} = -\frac{f'(\rho)}{g} ,
\label{3-96}\\
	H^5_\varphi (\rho) &=& \epsilon_{\varphi \rho z} F^{5\rho z} =
	- \frac{v'(\rho)}{g}.
\label{3-97}
\end{eqnarray}
We assume that 2-point Green functions can be approximately expressed as follows:
\begin{eqnarray}
	G^{ab \mu \nu} (y,x) &\approx& \Delta^{ab}
	\mathcal B^\mu \mathcal B^\nu ,\quad  a, b = 2, 5, 7 ;
\label{3-100}\\
	C^{mn \mu \nu} &\approx& \delta^{mn} \mathcal A^\mu \mathcal A^\nu ,
	\quad  m, n = 1, 3, 4, 6, 8 ,
\label{3-110}
\end{eqnarray}
where $\mathcal B_\mu \mathcal B^\mu$ is a constant and
\begin{eqnarray}
	\Delta^{ab} &=& \begin{pmatrix}
		\delta_1	& 0			&  0 	\\
		0			& \delta_2	&  0	\\
		0			& 0			&  \delta_3
	\end{pmatrix} ,
\label{3-120}\\
	\Delta^{mn} &=& \begin{pmatrix}
		\Delta_1	& 0			&  0 & 0 & 0	\\
		0			& \Delta_2	&  0 & 0 & 0	\\
		0			& 0			&  \Delta_3 & 0 & 0 \\
		0	& 0			&  0 & \Delta_4 & 0	\\
		0			& 0	&  0 & 0 & \Delta_5	\\
	\end{pmatrix}.
\label{3-130}
\end{eqnarray}
We choose the vectors $\mathcal A^\mu$ and $\mathcal B^\mu$  in the form
\begin{eqnarray}
	\mathcal A^\mu &=& \left( 0, 0, \mathcal A_\rho, \frac{\mathcal A_\varphi}{\rho} \right) ,
\label{3-140}\\
	\mathcal B^\mu &=& \left( 0, 0, \mathcal B_\rho, \frac{\mathcal B_\varphi}{\rho} \right).
\label{3-150}
\end{eqnarray}
With such a choice of the vectors $A^a_\mu$, $\mathcal B^\mu$, and $\mathcal A^\mu$, we have
\begin{eqnarray}
	\left( \mu^2 \right)^{1b t\nu} A^b_\nu =
	\frac{g^2}{4} \left( \mathcal B_\rho^2 + \mathcal B_\varphi^2 \right)
	\left( \delta_5 + \delta_7 \right) A^2_t &=& \mu_1^2 A^2_t,
\label{3-160}\\
	\left( \mu^2 \right)^{2b z\nu} A^b_\nu =
	- \frac{g^2}{4} \left( \mathcal B_\rho^2 + \mathcal B_\varphi^2 \right)
	\left( \delta_2 + \delta_7 \right) A^5_z &=& - \mu_2^2 A^5_z,
\label{3-170}\\
	\left( m^2 \right)^{1b t\nu} A^b_\nu = \frac{3}{4} g^2
	\left(
		\mathcal A_\rho^2 + \mathcal A_\varphi^2
	\right) \left(
		4 \Delta_1 + 4 \Delta_3 + \Delta_4 + \Delta_6
	\right) 	
	\phi^2 A^2_t &=& \frac{m^2_1}{4} \phi^2 A^2_t,
\label{3-180}\\
	\left( m^2 \right)^{2b z\nu} A^b_\nu = - \frac{3}{4} g^2
	\left(
		\mathcal A_\rho^2 + \mathcal A_\varphi^2
	\right) \left(
		\Delta_1 + \Delta_3 + 4 \Delta_4 + \Delta_6 + 3 \Delta_8
	\right)
	\phi^2 A^5_z &=& - \frac{m^2_2}{4} \phi^2 A^5_z,
\label{3-190}\\
	\left( m^2_\phi \right)^{ab \mu \nu} A^a_\nu A^b_\mu =
	\frac{g^2}{4} \left[
		\frac{4 \Delta_1 + 4 \Delta_3 + \Delta_4 + \Delta_6}
		{\Delta_1 + \Delta_3 + \Delta_4 + \Delta_6 + \Delta_8} \left( A^2_t \right)^2 -
		\frac{\Delta_1 + \Delta_3 + 4 \Delta_4 + \Delta_6 + 3 \Delta_8 }
		{\Delta_1 + \Delta_3 + \Delta_4 + \Delta_6 + \Delta_8} \left( A^5_z \right)^2
	\right]
	&=&
\nonumber \\
	\frac{g^2}{4} \left[
		m^2_{1,\phi} \left( A^2_t \right)^2 - m^2_{2,\phi} \left( A^5_z \right)^2
	\right] &&.
\label{3-200}
\end{eqnarray}
Let us consider the simplest case with
\begin{equation}
	\Delta_1 + \Delta_3 = \Delta_4 + \Delta_8.
\label{3-240}
\end{equation}
Then
\begin{eqnarray}
	m_2^1 = m_2^2 &=& m^2 = 3 g^2 \left(
		\mathcal A_\rho^2 + \mathcal A_\varphi^2
	\right) \left(
		5 \Delta_4 + \Delta_6 + 4 \Delta_8
	\right) ,
\label{3-245}\\
	m^2_{1,\phi} = m^2_{2,\phi} &=& m^2_{\phi} =
	\frac{5 \Delta_4 + \Delta_6 + 4 \Delta_8}{2 \Delta_4 + \Delta_6 + 2 \Delta_8}.
\label{3-250}
\end{eqnarray}
In order to have equations that are the Euler-Lagrange equations, from some Lagrangian we choose
\begin{equation}
	m^2 = m^2_\phi , \quad
	3 g^2 \left(
	\mathcal A_\rho^2 + \mathcal A_\varphi^2
	\right) = \frac{1}{2 \Delta_4 + \Delta_6 + 2 \Delta_8} .
\label{3-255}
\end{equation}
Substituting \eqref{3-90} and \eqref{3-160}-\eqref{3-200} into \eqref{3-10}  and \eqref{3-20} and taking into account \eqref{3-255}, we have
\begin{eqnarray}
	f'' + \frac{f'}{\rho} &=& f \left( - \frac{v^2}{4} + \frac{m^2}{4} \phi^2 - \mu^2_1 \right),
\label{3-210}\\
	v'' + \frac{v'}{\rho} &=& v \left( \frac{f^2}{4} + \frac{m^2}{4} \phi^2 - \mu^2_2 \right),
\label{3-220}\\
	\phi'' + \frac{\phi'}{\rho} &=& \phi \left[
		- m^2 \frac{f^2}{4} + m^2 \frac{v^2}{4}
		+ \lambda \left( \phi^2 - M^2 \right)
	\right].
\label{3-230}
\end{eqnarray}
The Lagrangian for this set of equations  is
\begin{equation}
	8 \pi \mathcal L = \frac{{f^\prime}^2}{2} - \frac{{v^\prime}^2}{2} -
	\frac{{\phi^\prime}^2}{2} + \frac{f^2 v^2}{8} + m^2 \frac{f^2 \phi^2}{8} -
	m^2 \frac{v^2 \phi^2}{8} - \frac{\mu_1^2}{2} f^2 + \frac{\mu_2^2}{2} v^2 -
	\frac{\lambda}{4} \left(
	\phi^2 - M^2
	\right)^2 .
\label{3-235}
\end{equation}
After introducing the dimensionless parameters $x = \rho \phi_0$,
$\tilde \phi = m \phi/2 \phi_0$,
$
\tilde \lambda = 4 \lambda / m^4
$, $\tilde M = m M/ 2\phi_0$, $\tilde f = f/(2 \phi_0)$ and $\tilde v = v/(2 \phi_0)$,
$\tilde \mu_{1,2} = \mu_{1,2}/\phi_0$, we have the following set of equations
\begin{eqnarray}
	\tilde f'' + \frac{\tilde f'}{\rho} &=& \tilde f \left(
		- \tilde v^2 + \tilde \phi^2 - \tilde \mu^2_1
	\right),
\label{3-260}\\
	\tilde v'' + \frac{\tilde v'}{\rho} &=& v \left(
		\tilde f^2 + \tilde \phi^2 - \tilde \mu^2_2
	\right),
\label{3-270}\\
	\tilde \phi'' + \frac{\tilde \phi'}{\rho} &=& m^2 \tilde \phi \left[
		- \tilde f^2 + \tilde v^2
	+ \tilde \lambda \left( \tilde \phi^2 - \tilde M^2 \right)
	\right],
\label{3-280}
\end{eqnarray}
where the prime denotes  differentiation with respect to the dimensionless radius $x$.

\subsection{Numerical solution}
\label{numSol}

\begin{figure}[h!]
	\begin{minipage}[ht]{.45\linewidth}
		\begin{center}
			\fbox{
				\includegraphics[width=.9\linewidth]{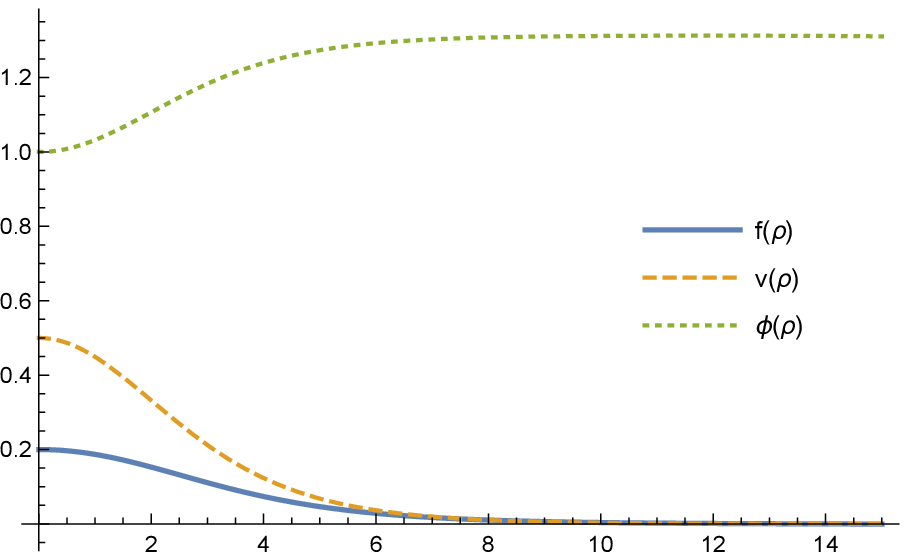}
			}
		\end{center}
		\caption{The functions $f(\rho), v(\rho), \tilde \phi(\rho)$.
			The solid curve is $f(\rho)$,
			the dashed curve  is $v(\rho)$,
			the dotted curve is $\tilde \phi(\rho)$.
			$\mu_1 \approx 1.2325683, \mu_2 \approx 1.180660003, M \approx 1.3137067$, $\tilde \lambda = 0.1$.
		}
		\label{potentials}
	\end{minipage}\hfill
	\begin{minipage}[ht]{.45\linewidth}
		\begin{center}
			\fbox{
				\includegraphics[width=.9\linewidth]{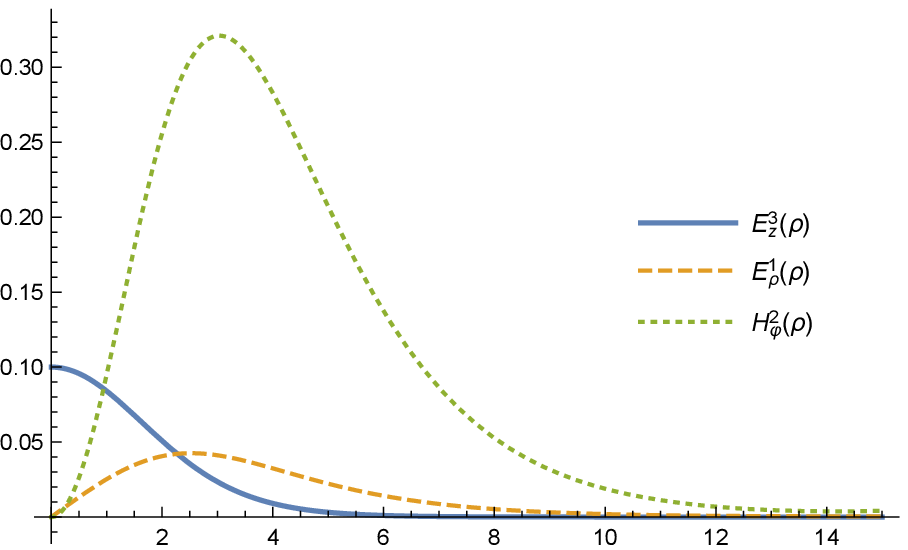}
			}
		\end{center}
		\caption{The chromoelectric and chromomagnetic fields
			$E^3_z(\rho), E^1_\rho(\rho), H^2_\varphi (\rho)$.
			The solid curve is $E^3_z(\rho)$,
			the dashed curve  is $E^1_\rho(\rho)$,
			the dotted curve is $H^2_\varphi (\rho)$
		}
		\label{fields}
	\end{minipage}
\end{figure}

In this subsection we want to present a numerical solution of the equations \eqref{3-260}-\eqref{3-280} which
are solved as a nonlinear eigenvalue problem with the eigenvalues $\mu_{1,2}, \tilde M$ and the eigenfunctions $\tilde f, \tilde v, \tilde \phi$. The boundary conditions are
\begin{equation}
\begin{split}
	f(0) &= 0.2, f'(0) = 0;
\\
	v(0) &= 0.5, v'(0) = 0;
\\
	\phi(0) &= 1.0, \phi'(0) = 0 .
\label{3-1-10}
\end{split}
\end{equation}
The results of calculations are presented in Figs. \ref{potentials} and  \ref{fields}. From Eqs. \eqref{3-260}-\eqref{3-280} we can obtain an asymptotic behavior of the functions
$f(\rho), v(x)$, and $\tilde \phi(\rho)$ in the form
\begin{eqnarray}
	\tilde f(\rho) &\approx& f_\infty \frac{e^{- x \sqrt{\tilde M^2 - \mu_1^2}}}{\sqrt{x}} ,
\label{3-1-20}\\
	\tilde v(\rho) &\approx& v_\infty \frac{e^{- x \sqrt{\tilde M^2 - \mu_2^2}}}{\sqrt{x}} ,
\label{3-1-30}\\
	\tilde \phi(\rho) &\approx& M -
	\phi_\infty \frac{e^{- x \sqrt{2 \tilde \lambda \tilde M^2}}}{\sqrt{x}},
\label{3-1-40}
\end{eqnarray}
where $f_\infty, v_\infty$, and $\phi_\infty$ are some constants.

Now we want to discuss to which quarks from  Fig \ref{protonModel1} the flux tube with the fields $E^7_z$ from \eqref{3-93} is attached. Analyzing the interaction term $\bar q A^B_\mu \lambda^B q$ from Lagrangian we see that: the flux tube with $E^7_z$ is attached to green, $q_g$, and blue,  $q_b$, quarks.

\section{Calculation of the angular momentum of gluon field}

\begin{figure}[h!]
\begin{minipage}[ht]{.45\linewidth}
	\begin{center}
		\fbox{
			\includegraphics[width=.9\linewidth]{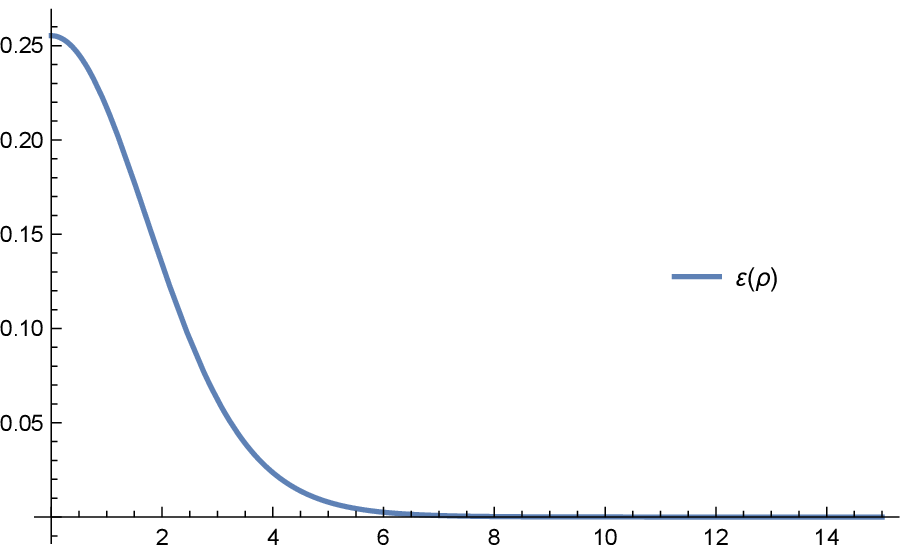}
		}
	\end{center}
	\caption{The profile of the energy density \eqref{5-30}.
	}
	\label{energyDensuty}
\end{minipage}
\begin{minipage}[ht]{.45\linewidth}
	\begin{center}
	\fbox{
		\includegraphics[width=.9\linewidth]{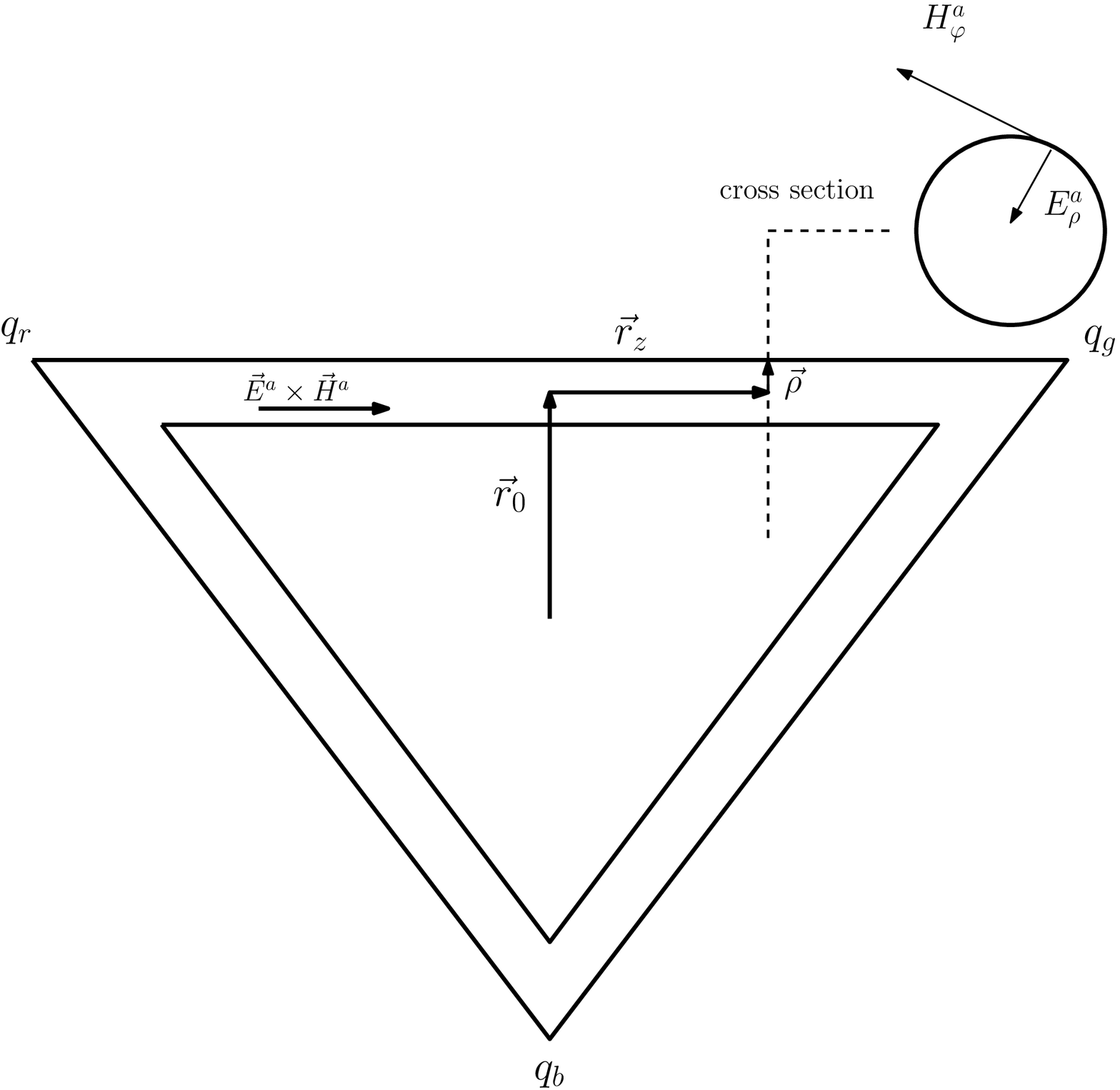}
	}
	\end{center}
	\caption{Longitudinal and cross sections of the flux tube with the directions of the fields.
		}
\label{mz}
\end{minipage}\hfill
\end{figure}

In this section we consider 
$A^2_{t \leftrightarrow z}, A^5_{z \leftrightarrow t}$ possible mechanism for the emergence of angular momentum of gluon field in proton spin. In section \ref{inftube} we have obtained the flux tube solution for $A^2_t$, $A^5_z$ SU(2) gauge potential. As it is easy to see the solution with potentials $A^2_z$, $A^5_t$ exists also. Both solutions are identical with the accuracy of exchange either $t \leftrightarrow z$ or $2 \leftrightarrow 5$. That means that in quantum theory should be a quantum state $\left. \left. \right| p \right\rangle $ describing both solutions. The properties of this quantum state are 
\begin{eqnarray}
	\left\langle p \left| 
		\hat E^2_\rho \hat H^5_\varphi
	\right| p \right\rangle &\neq& 0 , 
\label{4-a-10}\\
	\left\langle p \left| 
		\hat E^5_\rho \hat H^2_\varphi
	\right| p \right\rangle &\neq& 0 , 
\label{4-a-20}\\
	\left\langle p \left| 
		\hat E^2_\rho \hat H^2_\varphi
	\right| p \right\rangle &\neq& 0 , 
\label{4-a-30}\\
	\left\langle p \left| 
		\hat E^5_\rho \hat H^5_\varphi
	\right| p \right\rangle &\neq& 0 , 
\label{4-a-40}
\end{eqnarray}
The relations \eqref{4-a-30} and \eqref{4-a-40} means that there is a complicated quantum interplay between $A^2_t, A^5_z$ and $A^5_t, A^2_z$ solutions. The expectation value of the angular momentum of the gluon field for one flux tube is 
\begin{equation}
\begin{split}
	\hat{\vec M} = & \frac{1}{4 \pi c} 
	\left\langle p \left| 
		\int dV \left[
			\vec r \times \left[ \hat{\vec E}^a \times \hat{\vec H}^a \right]
		\right] 
	\right| p \right\rangle = 
\\
	& \frac{1}{4 \pi c} 
			\int dV \left[
				\vec r \times \left[ 
					\left\langle p \left| 
						\hat{\vec E}^2 \times \hat{\vec H}^2
					\right| p \right\rangle 	
				\right]
			\right] + 
	\frac{1}{4 \pi c} 
				\int dV \left[
					\vec r \times \left[ 
						\left\langle p \left| 
							\hat{\vec E}^5 \times \hat{\vec H}^5
						\right| p \right\rangle 	
					\right]
				\right] .
\end{split}
\label{4-a-50}
\end{equation}
For the calculation of \eqref{4-a-50} we approximate the flux tube as a finite piece of the infinite flux tube found in Section \ref{numSol}. Then the $M_\perp$  component of the angular momentum of the gluon fields in the flux tube, which is perpendicular to the triangle created by three quarks, will be
\begin{equation}
	L_G = M_{\perp, g} = 
	\left(
		\frac{24 \pi}{{g^\prime}^2} \tilde l \tilde r_0 \int \limits_0^\infty
		x \tilde f^\prime \tilde v^\prime d x
	\right) \frac{\hbar}{2} =
	\left(
		\frac{24 \pi}{{g^\prime}^2} l r_0 \phi_0^2 \tilde M_\perp
	\right) \frac{\hbar}{2} 
\label{4-a-60}
\end{equation}
here $\vec r = \vec r_0 + \vec r_z + \vec \rho$, see Fig. \ref{mz}; $\vec r_0$ is the distance from the proton center to the flux tube center; $\vec r_z$ is the distance from the flux tube center to the cross section of the tube; 
$\vec \rho$ is the distance from the flux tube axis to the point where 
$
\left[
	\vec r \times \left[ \vec E^a
		\times \vec H^a
	\right] 
\right]
$ is calculated; the transversal color electric and magnetic fields
$
E^{2,5}_\rho = -\frac{f'(\rho)}{g}
$,
$
H^{2,5}_\varphi = - \frac{v'(\rho)}{g}
$
are calculated in Eqs.~\eqref{3-96} and \eqref{3-97}; 
${g^\prime}^2 = 4 \pi c \hbar g^2$ is the dimensionless coupling constant;
$\tilde l = l \phi_0$ is dimensionless lengths of the tube; $\tilde r_0 = r_0 \phi_0$ where $r_0 \approx r_p$, $r_p$ is the proton radius; 
$\tilde M_\perp \approx 0.043$ can be calculated by using the numerical solution obtained in Section \ref{numSol}; we take into account that we have two equal terms in \eqref{4-a-50}, and three flux tubes in the proton; 
$d V = 2 \pi l \rho d \rho$. One way to calculate  $\phi_0$ is to calculate the mass of the gluon fields in the proton. One can note that similar calculations for meson, consisting from quark and antiquark with one flux tube between them, gives us $L_G = 0$ because $r_0 = 0$. For the rough estimation of $L_G$ we have: $\tilde l, \tilde r_0, g^\prime \approx 1$ and 
$M_{\perp, g}/(\hbar/2) \approx 4 \%$.

\section{Calculation of the mass of gluon fields in a proton}

Calculation of the proton mass of gluon fields in this approach faces the following problem: The energy density for the non-Abelian fields is
\begin{equation}
	\varepsilon = \frac{1}{8 \pi} \left(
		\left\langle \hat E^B_i \hat E^{B i} \right\rangle +
		\left\langle \hat H^B_i \hat H^{B i} \right\rangle
	\right),
\label{5-10}
\end{equation}
where $E^B_i$ and $H^B_i$ are non-Abelian electric and magnetic fields. The point is that in our approach (for details see Ref. \cite{Dzhunushaliev:2016svj})
\begin{equation}
	\hat A^a_\mu = A^a_\mu + i \delta \hat A^a_\mu .
\label{5-20}
\end{equation}
Here $a = 2,5,7$; $A^a_\mu \in SU(2) \subset SU(3)$;
$A^a_\mu = \left\langle \hat A^a_\mu \right\rangle $; $\delta \hat A^a_\mu$ are quantum fluctuations. In such a case the linear energy density will be
\begin{equation}
	8 \pi g^2 \epsilon(\rho) = \frac{{f^\prime}^2}{2} + \frac{{v^\prime}^2}{2} +
	\frac{{\phi^\prime}^2}{2} + \frac{f^2 v^2}{8} + m^2 \frac{f^2 \phi^2}{8} +
	m^2 \frac{v^2 \phi^2}{8} + \frac{\mu_1^2}{2} f^2 - \frac{\mu_2^2}{2} v^2 +
	\frac{\lambda}{4} \left(
		\phi^2 - M^2
	\right)^2 .
\label{5-30}
\end{equation}
As one sees from this expression,
there is negative term $- \frac{\mu_2^2}{2} v^2$ coming from \eqref{5-20}. Evidently, this is a consequence of the presence of the imaginary factor $i$ in front of
$\delta \hat A^a_\mu$ in \eqref{5-20}. To avoid this problem, we suggest the following approach: let us write the energy density as
\begin{equation}
	\varepsilon = \frac{1}{8 \pi} \left(
		\left\langle \left( \hat E^B_i \right)^\dagger \hat E^{B i} \right\rangle +
		\left\langle \left( \hat H^B_i \right)^\dagger \hat H^{B i} \right\rangle
	\right),
\label{5-40}
\end{equation}
where $\left( \hat E^B_i \right)^\dagger$ is the Hermitian conjugate operator. In this case the linear energy density will be
\begin{equation}
	8 \pi \epsilon(\rho) = \frac{{f^\prime}^2}{2} + \frac{{v^\prime}^2}{2} +
		\frac{{\phi^\prime}^2}{2} + \frac{f^2 v^2}{8} + m^2 \frac{f^2 \phi^2}{8} +
		m^2 \frac{v^2 \phi^2}{8} + \frac{\mu_1^2}{2} f^2 + \frac{\mu_2^2}{2} v^2 +
		\frac{\lambda}{4} \left(
		\phi^2 - M^2
	\right)^2 .
\label{5-50}
\end{equation}
Then the mass of the gluon fields in a proton is
\begin{equation}
\begin{split}
	m_g = & \frac{3}{4} \frac{l}{c^2 g^2} \int \limits_0^\infty
	\rho \varepsilon (\rho) d \rho =
\\
	&
	\frac{3 l \phi_0^2}{c^2 g^2} \int \limits_0^\infty
	x dx \left[
		\frac{{\tilde {f^\prime}}^2}{2} + \frac{{\tilde {v^\prime}}^2}{2} +
		\frac{{\tilde {\phi^\prime}}^2}{2} + \frac{\tilde f^2 v^2}{8} + \frac{\tilde f^2 \tilde \phi^2}{8} +
		\frac{\tilde v^2 \tilde \phi^2}{8} + \frac{\tilde \mu_1^2}{2} \tilde f^2 +
		\frac{\tilde \mu_2^2}{2} \tilde v^2 +
		\frac{\tilde \lambda}{4} \left(
			\tilde \phi^2 - \tilde M^2
		\right)^2
	\right] =
\\
	&
	3 \left( \frac{\hbar }{c} l \phi_0^2\right)
	\frac{1}{{g^\prime}^2} \tilde m_g.
\label{5-60}
\end{split}
\end{equation}
Here $\tilde m_g \approx 0.85$ can be calculated by using the numerical solution obtained in Section \ref{numSol}. Using \eqref{5-60},
we can exclude $\phi_0$ from \eqref{4-a-60} and obtain a dimensionless relation between $M_\perp$ and $m_g$:
\begin{equation}
	\frac{1}{2 c r_0} \frac{M_{\perp, g}}{m_g} \approx \frac{\tilde M_\perp}{\tilde m_g} .
\label{5-70}
\end{equation}
The left-hand side contains quantities which can be in principle measured, and the right-hand side is predicted theoretically. This value depends on the parameters $f(0), v(0), \lambda, m_{1,2}$ and, varying these parameters, more realistic
values of the gluon field angular momentum can be obtained.

\section{The contribution to proton spin from rotating quarks and tubes}

Besides the angular momentum of crossed colour electric and magnetic fields, the proton spin may have a contribution from the rotating quarks and flux tubes connecting quarks. We can estimate such contribution in the following way 
\begin{equation}
	M_{\perp, r} \approx \left( m_q + m_g \right) v r_0 
\label{6-10}
\end{equation}
here $v$ is the velocity of quarks and flux tubes connecting quarks. The total angular momentum should be 
$S_q + M_{\perp, g} + M_{\perp, r} = \hbar/2$ where $S_q$ is the contribution from quarks spin. After some algebraic manipulations we have 
\begin{equation}
	\frac{S_q}{m_p c r_0} + \frac{m_q}{m_p} \frac{v}{c} + \frac{m_g}{m_p} \left( 
		2 \frac{\tilde M_\perp}{\tilde m_g} + \frac{v}{c}
	\right) \approx 10^{-1}
\label{6-20}
\end{equation}
here $m_q$ is the quark masses and we took into account \eqref{2-10}.

\section{Discussion and conclusions}

Thus, we have approximately calculated the contribution of crossed gluon fields to the proton spin. The calculations are based on  Heisenberg's quantization technique. We have considered a proton as an object constructed from three quarks connected by three flux tubes. Using Heisenberg's quantization approach,
we have shown that the flux tubes contain color longitudinal and transversal electric and magnetic fields with different color indexes. This does not allow us to have a nonzero angular momentum density created by these transversal fields. But if we consider a quantum state where some interplay between these possibilities is possible then we will have transversal electric and magnetic fields with the same color index. It allows us to have the nonzero angular momentum density.

We have calculated the component of the angular momentum of crossed colour fields perpendicular to the flux tubes. In order to estimate this value, we have calculated the mass of the gluon fields. Having this mass, we could determine one unknown parameter $\phi_0$ which is the value of the dispersion of quantum fluctuations of the coset fields at the proton centre. Then we have estimated the contribution to proton spin coming from rotating quarks and flux tubes connecting quarks. 

The calculations presented here depend on the flux tube solution found in Section \ref{numSol}. The solution depends on the parameters $m_\phi, \tilde \lambda, v(0), \phi(0)$.
These parameters cannot be determined in the two-equations approximation  but only in the next approximation: three-, four-, and so on equations approximation.

\section*{Acknowledgements}

This work was supported by Grant $\Phi.0755$  in fundamental research in natural sciences by the MES of RK. I am very grateful to V. Folomeev for fruitful discussions and comments.

\end{document}